# SUBSTRATE INTEGRATED WAVEGUIDE POWER DIVIDER, CIRCULATOR AND COUPLER IN [10-15] GHZ BAND

Bouchra Rahali and Mohammed Feham
STIC Laboratory, University of Tlemcen, Tlemcen 13000, Algeria
b_rahali@hotmail.fr
m.feham@mail.univ-tlemcen.dz

## ABSTRACT

*The Substrate Integrated Waveguide (SIW) technology is an attractive approach for the design of high performance microwave and millimeter wave components, as it combines the advantages of planar technology, such as low fabrication costs, with the low loss inherent to the waveguide solution. In this study, a substrate integrated waveguide power divider, circulator and coupler are conceived and optimized in [10-15] GHz band by Ansoft HFSS code. Thus, results of this modeling are presented, discussed and allow to integrate these devices in planar circuits.*

## 1. INTRODUCTION

SIW technology [1][2] is interesting for the benefits of rectangular waveguides while maintaining in planar profiles. Easy integration and a high quality factor are interesting characteristics of the rectangular waveguide in the technology SIW (RSIW). many SIW components such as bends [3], filters [4], couplers [5], duplexers [6], sixports [7], circulators [8] and phase shifters [9] has been studied. Figure 1 illustrate the RSIW which is designed from two rows of periodic metallic posts connected to higher and lower planes mass of dielectric substrate. Figure 2 and 3 shows the similarity of the geometry and the distribution of the electric field between (RSIW) and the equivalent rectangular wave guide [2] [3]. In this paper, [10-15] GHz band RSIW components are proposed and optimized. They are essential for many microwave and millimeter-wave integrated circuits and telecommunication systems.

## 2. FUNDAMENTAL RSIW CHARACTERISTICS

For designing (RSIW) (Figure 1) many physical parameters are necessary as d the diameter of holes stems, p the spacing between the holes and $W_{SIW}$ spacing between the two rows of holes. Between the two metal planes of the dielectric substrate two rows of holes are drilled and metalized permitting propagation for all modes $TE_{n0}$ [5]. The lines of current along the lateral walls of the RSIW are vertical, the fundamental $TE_{10}$ mode may be propagated. Electrical performance of RSIW and a conventional rectangular waveguide filled with dielectric of width $W_{eq}$ [4] are similar.

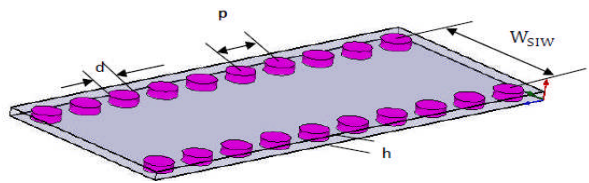

Figure 1. Rectangular wave guide integrated into a substrate RSIW

For obtaining the same characteristics of the fundamental mode propagating in the RSIW (Figure 2) having the same height and the same dielectric, empirical equations [2] were derived in order to determine the width of the equivalent rectangular wave guide

$$W_{eq} = W_{SIW} - \frac{d^2}{0.95\,p} \quad (1)$$

$$p < \frac{\lambda_0}{2} \sqrt{\varepsilon_r} \quad (2)$$

$$p < 4\,d \quad (3)$$

$$\lambda_0 = \frac{c}{f}$$

Where $\lambda_0$ is the space wavelength.

Period p must be low for reducing leakage losses between adjoining cylinders. We examined through this study, the RSIW [10-15] GHz from a conventional waveguide [10], the feature parameters are outlined in Table 1. We deduce the parameters of RSIW and the equivalent waveguide (Figure 2) Table 1 from the approach cited in [2].

Table 1

| Classic wave guide | Equivalent wave guide | RSIW |
|---|---|---|
| WR75 a=18.35mm, b=9.175mm, $\varepsilon_r=1$ | h=0.8mm, $\varepsilon_r=2.2$ $W_{eq}=10.73$mm | h=0.8mm, $\varepsilon_r=2.2$, d=0.5mm, p=1mm, $W_{SIW}=11$mm |

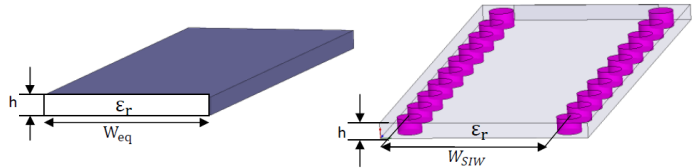

Figure 2. Equivalent rectangular waveguide and RSIW

HFSS tool [11] based on the finite element method (FEM) allows to optimize initial values $W_{SIW}$ given by (1), (2) and (3). It also helps give the scatter diagram and layout of the cartography of the electromagnetic field of the $TE_{10}$ mode. Through Figure 3 we note similarity between electromagnetic field distribution of $TE_{10}$ mode guided in equivalent rectangular waveguide and RSIW .

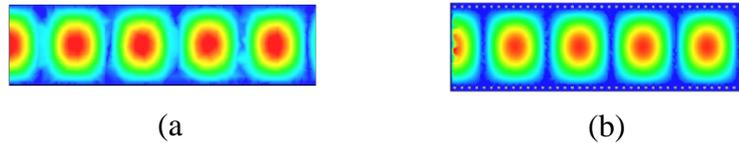

(a)          (b)

Figure 3. Electric field distribution of the $TE_{10}$ mode in the equivalent rectangular waveguide (a) and RSIW (b) at the frequency f = 12 GHz

Also Figure 4 shows, between these two waveguides, the coherence of the dispersion characteristics . It is worth noting that the similarity of propagation is valid for all modes $TE_{n0}$.

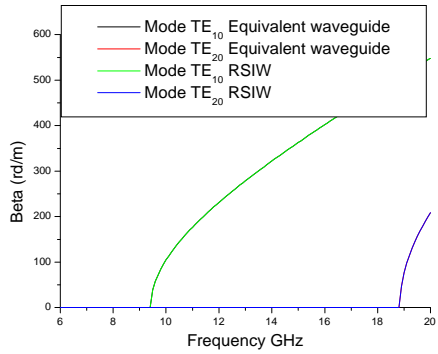

Figure 4. Dispersion characteristics

## 2. RSIW-MICROSTRIP TAPERED TRANSITION

For interconnect RSIW to the planar transmission lines the microstrip transition taper [12] is employed .A tapered section is used to match the impedance between a 50 Ω microstrip line in which the dominant mode is quasi-TEM and $TE_{10}$ mode of the RSIW, their electric field distributions are approximate in the profile of the structure.

From several formulas given [13] initial parameters $W_T$ and $L_T$ are determined and optimized with HFSS [11] Table 2. Figure 5, 6 and 7 shows the proposed transitions of coplanar taper of dimensions $L_T$ , $W_T$ to RSIW and mentioned the results of the RSIW simulation without transition and with microstrip line to RSIW.

Table 2

| | |
|---|---|
| $L_T$ | 2.1mm |
| $W_T$ | 3.81mm |
| $W_{mst}$ | 2.41mm |
| L | 40.016mm |

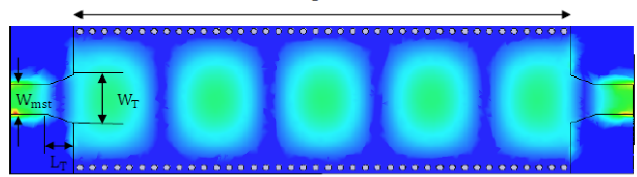

Figure 5. Electric field distribution of $TE_{10}$ mode at f = 12 GHz in the matched RSIW

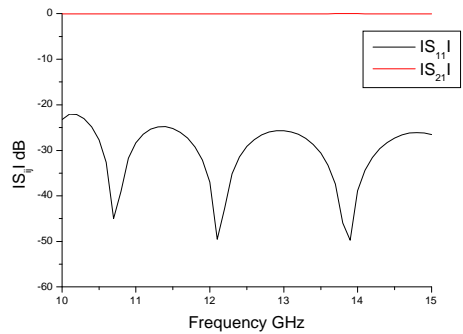

Figure 6. Transmission coefficients $S_{21}$ and reflection $S_{11}$ of the RSIW

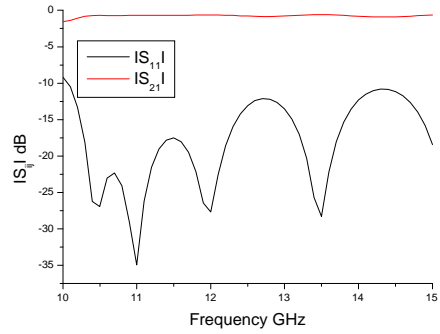

Figure 7. Transmission coefficients $S_{21}$ and reflection $S_{11}$
of the matched RSIW with taper

The reflection coefficient $S_{11}$ is less than -15 dB over 18.8% of the frequency band and the transmission coefficient $S_{21}$ is around - 0.63 dB across the entire band (Figure 7). On the same substrate and without any mechanical assembly [14] [15], this transition [13] allows the conception of a completely integrated planar circuit.

## 3. DESIGN OF RSIW PASSIVE DEVICES

### 2.1. SIW Power divider

There are two kinds of power dividers, T and Y [14] [15] [16] which are required in many applications to deliver copies of a signal in a system. This study is interested in the three ports power dividers where the half power (-3 dB) of an input signal is supplied to each of the two output ports.

From Table 1 and 2, we analyzed power divider (Figure 8), designed on three identical RSIW connected to form a T with L=14.5mm. So as to minimize reflection losses at the input port an inductive metal cylinder of radius r and position xp is inserted in this junction. We fix the radius r to the corresponding available practical value of diameter drills, and then change xp to get acceptable limit of reflection losses below -15dB. To each port a taper transition is added in order to integrate this component directly into a microstrip circuit. The input wave (port 1) is distributed equally into two parts which output to port 2 and port 3. Its matrix [S] (4):

$$[S] = \begin{bmatrix} S_{11} & S_{12} & S_{13} \\ S_{21} & S_{22} & S_{23} \\ S_{31} & S_{32} & S_{33} \end{bmatrix} \quad (4)$$

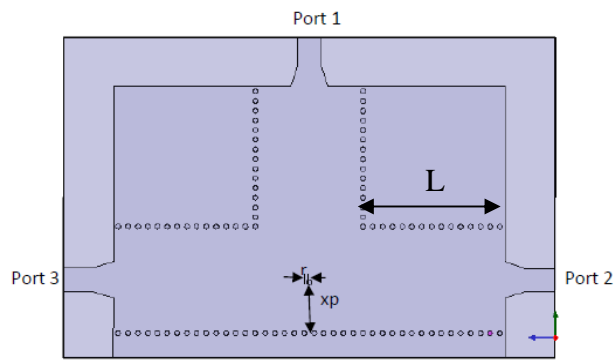

Figure 8. RSIW power divider

The distribution of the electric field of the $TE_{10}$ mode in the band [10-15] GHz, transmission coefficients $S_{21}$, $S_{31}$ and reflection coefficient $S_{11}$ of the power divider RSIW are related through Figures 9 and 10 respectively.

$S_{11}$ is less than -15 dB between 10.87 GHz and 12.44 GHz which is more than 17.3 % of the bandwidth. The optimal values of the inductive cylinder are r =0.254mm, xp=5.25mm. Transmission coefficients $S_{21}$ and $S_{31}$ fluctuate between -3.17 dB and - 3.34 dB being very acceptable levels.

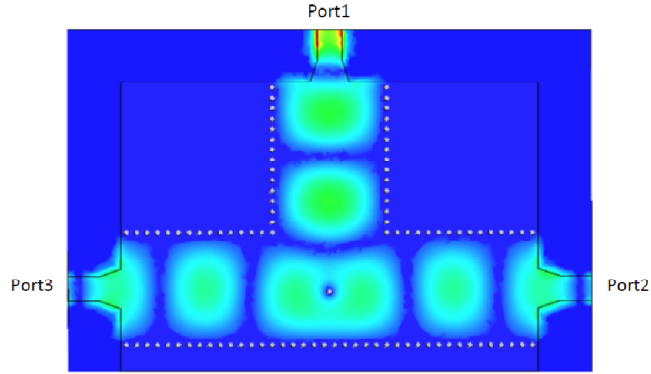

Figure 9. Electric field distribution of the $TE_{10}$ mode at f = 12 GHz in the RSIW power divider with inductive cylinder

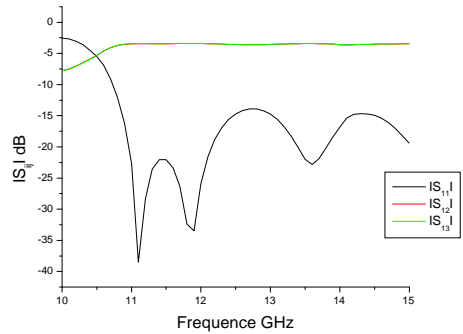

Figure 10. Parameters $S_{ij}$ in the RSIW power divider with inductive cylinder

## 2.2. SIW Circulator

Several types of solutions are common and recommended to provide protection microwave sources. The circulator waveguide technology [8] is the best solution in high power. Around a central body of ferrite (nickel materials and lithium ferrite) [8] [9], to which is applied a vertical magnetic field, we analyzed circulator (Figure 11), designed on three identical RSIW L=9.016mm and connected to have three access separate of 120° from each other. Indeed, incoming wave from port 1, 2 or 3 cannot out by the access 2, 3 or 1, respectively, conferring to the circulator the property of non-reciprocity.

Its ideal matrix [S] would be as follows:

$$[S] = \begin{bmatrix} 0 & 0 & e^{j\varphi} \\ e^{j\varphi} & 0 & 0 \\ 0 & e^{j\varphi} & 0 \end{bmatrix} \quad (5)$$

φ is the phase shift corresponding to transmission of the access signal to the next access. The non-reciprocity of the device shown by the non-symmetry of the matrix [S] is the main point explaining that this function can be used in many applications in telecommunications.

The saturation magnetization of ferrite material is [9] 4πMs = 1250 Gauss. Its relative dielectric constant is 13.7 and a radius $R_f$ calculated by [8].

$$R_f = \frac{1.84\ c}{\omega_0 \sqrt{\varepsilon_f}} \qquad (6)$$

c and $\omega_0$ are respectively the velocity of light in the free space and the operation frequency [7]. The ferrite radius and height are $R_f$=2.3mm $h_f$=0.8mm.

Through Figure 12 and 13 the distribution of the electric field of the $TE_{10}$ mode in RSIW circulator and the frequency response of RSIW circulator, transmission coefficients $S_{21}$, reflection coefficients $S_{11}$ and isolation coefficients $S_{31}$ are reported. In this frequency band the reflection loss $S_{11}$ below-15 dB take more than 10.75% of the bandwidth , the insertion loss $S_{21}$ is around of -0.43 dB , while the maximum of the isolation $S_{31}$ is -43.85 dB. At frequency of 12.5 GHz, the two figures 12 and 13 confirm the property of the circulation component [10].

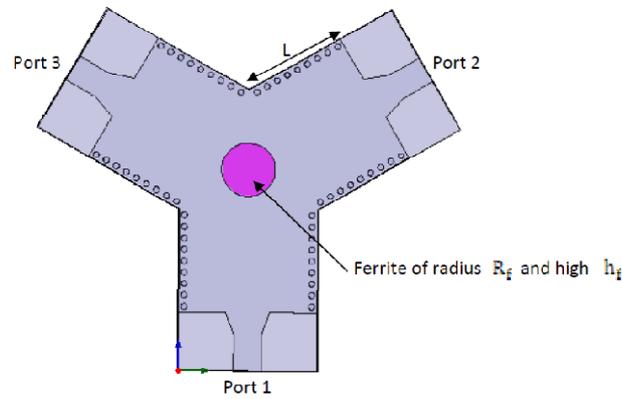

Figure 11. RSIW circulator

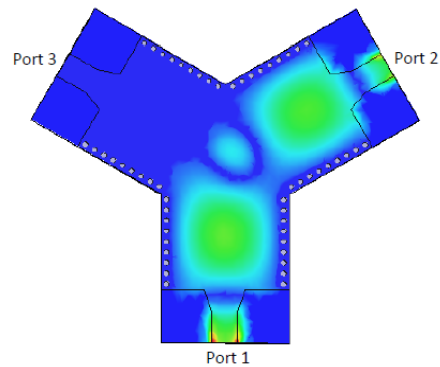

Figure 12. Electric field distribution of the $TE_{10}$ mode
of the RSIW circulator at f = 12.5 GHz

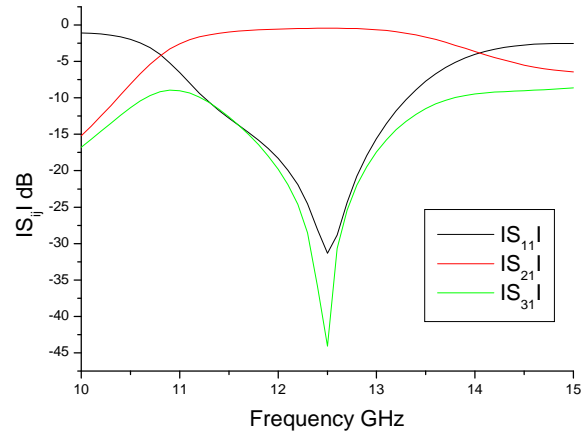

Figure 13. Parameters $S_{ij}$ of RSIW circulator

## 2.2. SIW Coupler

For routing, dividing and combining the signals in many microwave system the directional couplers [5] have been widely used as key component. The RSIW directional couplers [5] [17] are extensively investigated Figure 14. The RSIW directional coupler (-3dB) is conceived by two RSIW with a common wall on which an aperture realize the coupling between these two guides. With the same parameters presented in tables 1 and 2, this coupler was designed by using metal rods of square section [18]. From an even/odd mode analysis, where the even mode is the $TE_{10}$ and the odd mode is the $TE_{20}$ the geometry of the coupler is determined [15][19] with the phase difference $\Delta\varphi$ (7)

$$\Delta\varphi = (\beta_1 - \beta_2)W_{ap} \qquad (7)$$

$\beta_1$ and $\beta_2$ are the propagation constants of the $TE_{10}$ and $TE_{20}$ modes, respectively. The condition $\Delta\varphi = \pi/2$ needs to be satisfied in the frequency band. To each port a taper transition is added in order to integrate this component directly into a microstrip circuit. The matrix [S] of a symmetric coupler (-3 dB), adapted to its access, is given by equation (8).

$$[S] = \frac{1}{\sqrt{2}} \begin{bmatrix} 0 & 1 & j & 0 \\ 1 & 0 & 0 & j \\ j & 0 & 0 & 1 \\ 0 & j & 1 & 0 \end{bmatrix} \qquad (8)$$

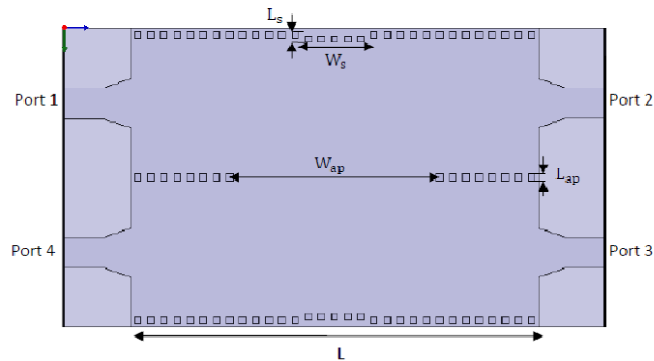

Figure 14. RSIW coupler

To achieve wide-band performance the coupler parameters are optimized using simulation under HFSS [11]. The final dimensions of the RSIW coupler are presented in Table 3.

Table 3

| | |
|---|---|
| L | 31.016mm |
| $W_s$ | 6mm |
| $L_s$ | 0.3mm |
| $W_{ap}$ | 16mm |
| $L_{ap}$ | 0.5mm |

Electric field distribution of $TE_{10}$ mode and the reflection coefficients $S_{11}$, the transmission coefficients $S_{21}$, the coupling coefficient $S_{31}$ and the isolation coefficient $S_{41}$ are related in the [10-15] GHz band through Figures 15 and 16 . The levels of reflection and isolation are below -15 dB with more than 20.34% of the bandwidth, and which the insertion loss $S_{21}$ and coupling $S_{31}$ are around -3.09±0.7 dB. These results show clearly the directional coupler character in the [10-15] GHz band.

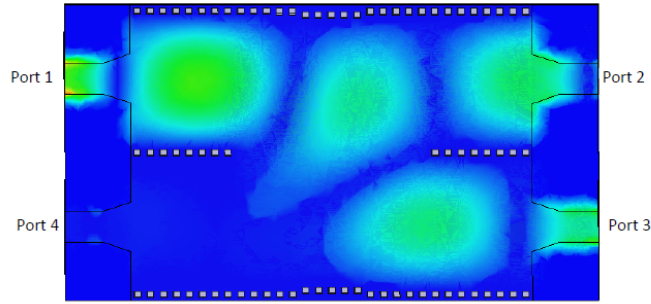

.

Figure 15. Electric field distribution of $TE_{10}$ mode for RSIW coupler at f=12 GHz

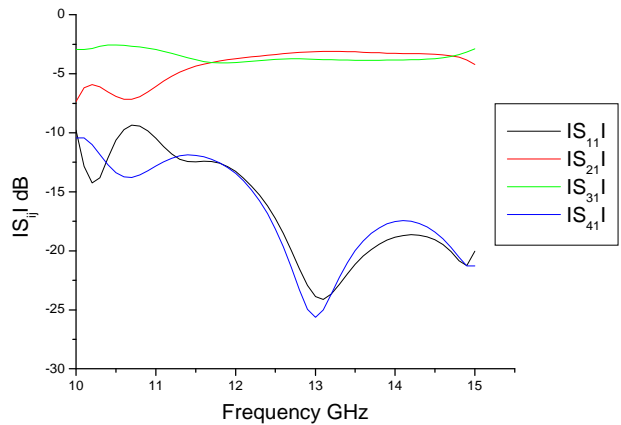

Figure 16. Frequency response of the RSIW directional coupler

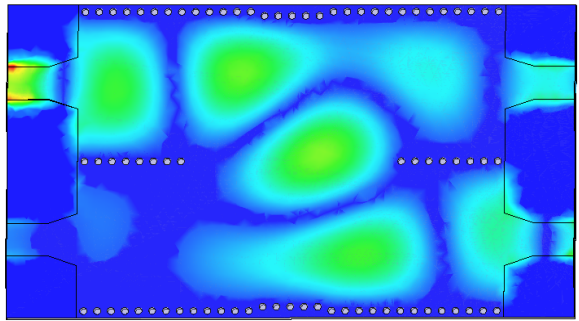

Figure 17. Electric field distribution of $TE_{10}$ mode for RSIW coupler at f=12 GHz

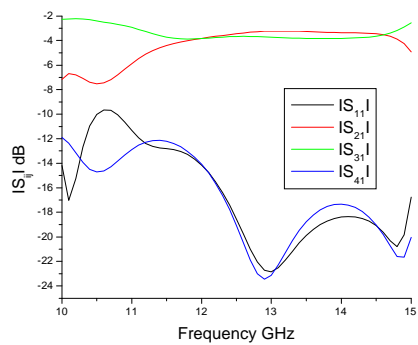

Figure 18. Frequency response of the RSIW directional coupler

## 3. CONCLUSIONS

In [10-15] GHz band substrate integrated waveguide passive components have been designed, using Ansoft HFSS code. Mainly our study focused on the RSIW three ports power dividers, the RSIW circulator and the RSIW coupler. They can be used as essential components in the conception of planar microwave and millimeter wave on the same substrate with microstrip. Through the results we see the good performance of these integrated devices conceived in a [10-15] GHz band.